\UseRawInputEncoding
\documentclass{WileyMSP-template}
\usepackage[english]{babel}

\usepackage[colorlinks=true, linkcolor=black, citecolor=red, linktocpage=false, breaklinks=false, urlcolor=black]{hyperref}

\usepackage{bigstrut}
\usepackage{multirow}
\usepackage{array}
\usepackage{tabularx}

\usepackage{graphicx}
\usepackage{amsmath}
\usepackage{mathtools}
\usepackage{siunitx}
\DeclareSIUnit\bar{bar}

\usepackage[super,compress]{cite}

\makeatletter
\renewcommand\@citess[1]{\textsuperscript{[#1]}}

\usepackage{letltxmacro}
\LetLtxMacro{\oldcite}{\cite}
\renewcommand{\cite}[1]{\mbox{\oldcite{#1}}} %verhindere, dass cite zeichen über 2 zeilen gehen

\begin{document}
%All text 
\justifying

\pagestyle{fancy}
\rhead{\includegraphics[width=2.5cm]{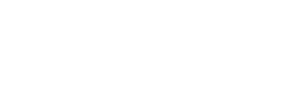}}

\title{Ferroelastic domain walls in BiFeO$_3$ as memristive networks}

\maketitle

% Author: Please give full first and last names for authors and include * after the name of all corresponding authors

\author{Jan Rieck*,}
\author{Davide Cipollini, }
\author{Mart Salverda, }
\author{Cynthia P. Quinteros, }
\author{Lambert R. B. Schomaker, }
\author{Beatriz Noheda*}

% Dedication

%\dedication{Optional dedication here. If no dedication is required, please leave blank}

% Affiliations: Please provide academic titles (Prof. or Dr.) for all authors where applicable, and include an institutional email address for all corresponding authors

\begin{affiliations}
J. Rieck, M. Salverda, Dr. C. P. Quinteros, Prof. Dr. B. Noheda\\
Zernike Institute for Advanced Materials, University of Groningen, Nijenborgh 4, 9747 AG Groningen, The Netherlands\\
Cognigron - Groningen Cognitive Systems and Materials Center, Nijenborgh 4, 9747 AG Groningen, The Netherlands\\

D. Cipollini, Prof. Dr. L. R. B. Schomaker \\
Bernoulli Institute for Mathematics, Computer Science and Artificial Intelligence, University of Groningen, 
Nijenborgh 9, 9747 AG Groningen, The Netherlands\\
Cognigron - Groningen Cognitive Systems and Materials Center, Nijenborgh 9, 9747 AG Groningen, The Netherlands\\

Cynthia P. Quinteros\\
ECyT-UNSAM, CONICET, Martin de Irigoyen 3100, B1650JKA, San Martín, Bs As, Argentina\\

Email Address: j.l.rieck@rug.nl
Email Address: b.noheda@rug.nl
%Email Address: d.cipollini@rug.nl
%Email Address: l.r.b.schomaker@rug.nl 

\end{affiliations}

% Keywords: Please provide a minimum of three and a maximum of seven keywords, separated by commas


% Abstract should be written in the present tense and impersonal style (i.e., avoid we), and be at most 200 words long

\begin{abstract}
    \noindent {Electronic conduction along individual domain walls (DWs) has been reported in BiFe$\text{O}_3$ (BFO) and other nominally insulating ferroelectrics. DWs in these materials separate regions of differently oriented electrical polarization (domains) and are just a few atoms wide, providing self-assembled nanometric conduction paths. In this work, it is shown that electronic transport is possible also from wall to wall through the dense network of as-grown DWs in BFO thin films. Electric field cycling at different points of the network, performed locally by conducting atomic force microscope (cAFM), induces resistive switching selectively at the DWs, both for vertical (single wall) and lateral (wall-to-wall) conduction. These findings are the first step towards investigating DWs as memristive networks for information processing and \textit{in-materio} computing.}
\end{abstract}

\section*{Introduction}

As our knowledge on memristive devices consolidates, the interest slowly moves towards the behaviour and functionality of networks of these devices. \cite{Adamatzky2014,Prezioso2015,DBLP:books/sp/19/CSA2019,Zegarac2019,Zhu2021} The resistive switching phenomena at the basis of memristive functionality \cite{Waser2009c,Waser2009d} originate from different underlying mechanisms (ion migration, redox reactions, ferroelectric switching, spin transfer torque, etc). \cite{Yang2013,Sillin2013,IelminiDaniele2016Rs:f,DBLP:books/sp/19/CSA2019,Minnai2017,Lanza2019,Milano2020} Memristive devices have been proposed as contenders for high-density, two-terminal, non-volatile random access (digital) memory.\cite{Sawa2008a,Kawahara2012,Liu2014,Schenk2020} In addition, their multiple resistance values bring them close to the behaviour of synapses (non-volatile variable resistance) and neurons (volatile variable resistance), offering them as the basic elements in neuromorphic computing applications \cite{Indiveri2013,DBLP:books/sp/19/CSA2019,Christensen2022}.
Nonetheless, the learning ability of the brain arises from a highly interconnected network of such elements in ways that are far from being understood, making the study of memristive networks highly relevant. \cite{Bak2001,Chialvo2010,DBLP:books/sp/19/CSA2019}

\par

Connecting memristive units together to form cross-bar arrays has been shown to allow extremely efficient Vector-Matrix Multiplication, putting forward memristive devices as synaptic weights for the implementation of Artificial Neural Networks in hardware, allowing the realization of analog resistive states.\cite{Jo2010a,Likharev2011,Prezioso2015,Merced-Grafals2016} Moreover, a network of memristors can effectively behave as a memristor with increased tunability in the on/off ratios, as well as in the threshold voltages\cite{Vourkas2015}, since the current flow not only depends on the history of the applied voltage, as in single memristors, but also heavily on the location of the input leads within the network.\cite{DiFrancesco2021a,Milano2022} It has also been shown that memristive networks are more robust to failure and variability than individual memristors, \cite{Choi2015,Hajto2019,Ernoult2019} which is of much importance, as the variability of memristive devices is the main issue in the way towards their implementation in hardware. In addition, a sufficiently large number of interconnected simple elements\textemdash such as memristors\textemdash is expected to display emergent behaviour \cite{Diaz-Alvarez2019,Chialvo2010,Dunham2021,Heywood2022}, which in the case of information processing, has been reported to allow complex learning functions with extreme energy savings \cite{Wang2014,Wang2020res,Christensen2022}. \par

In this work, networks of ferroelectric-ferroelastic domain walls (DWs), which are the boundaries between two domains (regions with differently-oriented electrical polarization), are investigated for their potential use as memristive networks. These DWs are one or two atoms wide \cite{Jia2008} and they self-assemble in ferroelectric materials to accommodate electrical and elastic boundary conditions. The density of DWs can be tuned by the choice of substrate and the system dimensions, such that the distance between domain walls can be as small as a few tens of nanometers in thin films \cite{Vlooswijk2007,Nesterov2013a,Feigl2014b}. Although ferroelectric materials are typically insulators, the DWs in some ferroelectrics have been reported to display enhanced conductivity compared to that in the domains \cite{Seidel2009a,Meier2012a,Catalan2012,Sluka2013a,Stolichnov2015,Kim2016a,Zhang2019}. Conductivity at DWs was first demonstrated by artificially switching selected areas of the sample using piezo-force microscopy (PFM) \cite{Ahn1997} and, subsequently, performing conducting atomic force microscopy (cAFM) maps around the newly created DWs \cite{Seidel2009a,Guyonnet2011}. However, it was also reported that as-grown ferroelastic domain walls that form during the growth process in BiFe$\text{O}_3$ (BFO) could display enhanced conductivity as well \cite{Chiu2011,Farokhipoor2011a,Farokhipoor2012}. Interestingly, DWs in different oxides have also been reported to be not only conductive\cite{Meier2012a,Ruff2017} but also memristive. \cite{Maksymovych2011,Lindgren2017,McConville2020,Chaudhary2020} Therefore, these materials could provide dense self-assembled memristive networks. \par

Ferroelastic DWs are formed during the growth process to release the epitaxial strain imposed by the substrate.\cite{Schilling2006a} Thus, unlike networks of metallic nanoparticles or nanowires \cite{Avizienis2012,Sandouk2015,Hochstetter2021}, or unlike artificially-created ferroelectric DWs,\cite{Gregg2022a} ferroelastic DW networks provide fixed conduction channels that are not easily moved, removed or created with an electric field. Therefore, the plasticity of the network is determined by the effect of ionic migration (driven by the strong strain gradients present around the ferroelastic DWs) on the electronic band bending \cite{Chiu2011,Farokhipoor2011a,Sluka2012,Wei2017a}, bringing some unique features. Despite their interest, previous works on self-assembled DWs mainly focus on the "vertical", out-of-plane (OOP), electrical response. In this paper, the possibility of obtaining "lateral", in-plane (IP), conduction through the DW network and, thus, to achieve charge flow parallel to the surface, along the walls, is investigated. First hints that this is, indeed, possible are presented. \par

\section*{Results}

\begin{figure}[htbp]
        \centering
		\includegraphics[draft=false,width=0.9\textwidth]{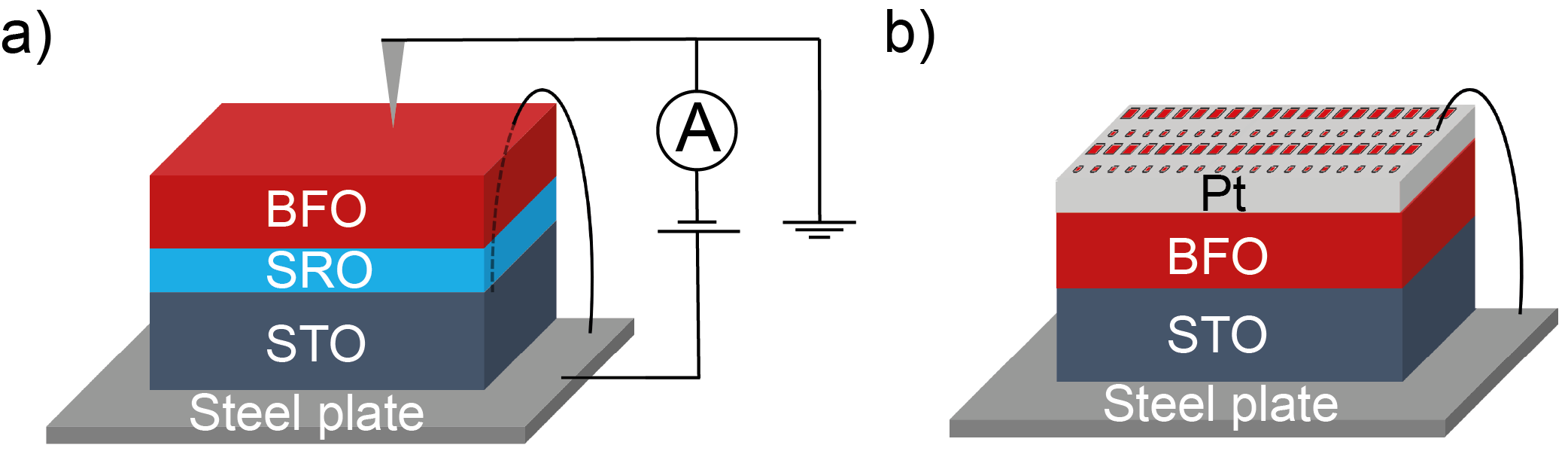}
	\caption[Vertical setup]{Experimental setup and layout of samples for \textbf{a)} vertical (OOP) and \textbf{b)} lateral (IP) conduction measurements. For the OOP conduction experiments, a SrRuO$_3$ (SRO) layer is deposited between the STO substrate and the BFO layer to serve as  bottom electrode. The SRO layer is electrically connected by means of wire-bonding. A Pt top electrode is patterned to leave square windows that enable access to the BFO surface.}
	\label{double_setup}
\end{figure}
\noindent

BFO thin films with a thickness of \SI{55}{\nano\meter} were deposited by pulsed laser deposition (PLD) on TiO$_2$-terminated (100) SrTiO$_3$ (STO) single crystal substrates. Two types of samples, with and without bottom electrode, have been fabricated. OOP transport measurements are performed on samples with SrRuO$_3$ (SRO) buffer layers acting as the bottom electrode. These samples are referred to as BFO/SRO/STO. For IP measurements, samples without bottom electrode are used. These are referred to as BFO/STO. The measurements are performed using the conductive tip of an AFM as top electrode (cAFM) in two different geometries, as shown in Figure \ref{double_setup}. More details of sample fabrication and measurement techniques are found in the Experimental Section.  \par

%the patterned Pt layer on top is used instead for laterally polarizing the film

\begin{figure}[htbp]
        \centering
		\includegraphics[draft=false,width=\textwidth]{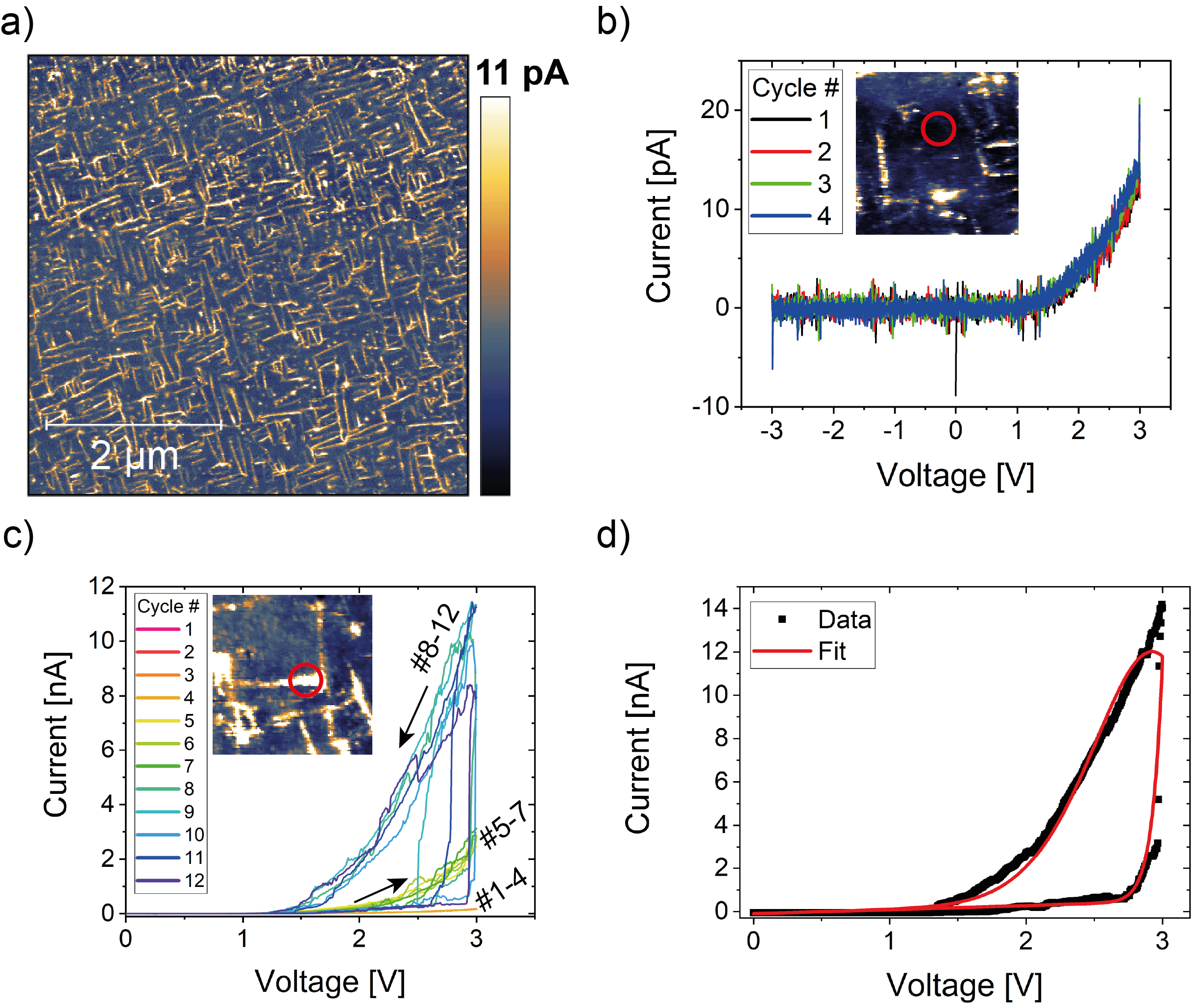}
	\caption[OOPimagesp]{Out-of-plane (OOP) domain wall conduction. \textbf{a)} OOP conduction map of a BFO/SRO/STO sample, obtained by cAFM with a \SI{3}{\volt} sample bias and a grounded tip. The as-grown ferroelastic/ferroelectric DWs show enhanced conduction compared to the domains. \textbf{b-c)} OOP \textit{I}-\textit{V} curves probed on the same BFO/SRO/STO sample for contacting a ferroelectric domain (\textbf{b)}) and an individual domain wall (\textbf{c)}). The respective probed locations are given by the red circles on the conduction maps shown as insets. \textbf{d)} Fit of an \textit{I}-\textit{V} loop measured in the same location as indicated in \textbf{c)} by a model based on an equivalent circuit combining a memristor and a diode.\cite{Yang2008}}
	\label{OOP}
\end{figure}

PFM images measured on a BFO/SRO/STO sample shown in Figure S1 (see Supporting Information) show agreement with previously reported data on BFO thin films: the as-grown BFO films are down-polarized, with four domain types present, and the DWs are of the 71$^\text{o}$ type, aligned in two orthogonal directions.\cite{Xu2005,Bea2007,Daumont2010,Farokhipoor2011} The OOP conduction map of the same BFO/SRO/STO sample is shown in Figure \ref{OOP} \textbf{a)}. The scans are performed with a sample bias of \SI{3}{\volt} on the bottom electrode, while the tip is grounded. The domain structure gives rise to a close-meshed, well-interconnected network of domain walls that are more conducting than the host material, in agreement with previous reports\cite{Farokhipoor2011a}.\par

\noindent

\newpage

In this OOP geometry, local current-voltage (\textit{I}-\textit{V}) characteristics can be obtained both inside the domains and at domain walls, by locally placing the conducting tip on selected positions at the sample surface and applying an alternating voltage signal. \textit{I}-\textit{V} curves measured inside a domain are shown in Figure \ref{OOP} \textbf{b)}. They were obtained applying a triangular wave with a frequency of \SI{1}{\hertz} and an amplitude of \SI{3}{\volt}. A diode-like behavior is observed: while no current response is measured for negative voltages (negative polarity at the bottom electrode), positive voltages induce a maximum current of \SI{20}{\pico\ampere}. No significant change is observed upon further cycling. The rectifying behaviour can be explained by the different work functions of the electrodes. The existence of different Schottky barriers at both interfaces, namely the CoCr alloy tip /BFO top interface and the BFO/SRO bottom interface, has been previously reported \cite{Farokhipoor2014a,Dedon2016}.\par

%\newpage
In Figure \ref{OOP} \textbf{c)}, \textit{I}-\textit{V} curves probing a highly conductive domain wall are shown. The same triangular voltage signal used in Figure \ref{OOP} \textbf{b)} was applied over a duration of 12 cycles. Similar to the response inside the domains, rectifying behaviour is observed with no current response for negative voltages, while for positive voltages values of up to \SI{10}{\nano\ampere}, that is three orders of magnitude larger than in the domains, can be measured. In this case, the \textit{I}-\textit{V} characteristics evolve with electric field cycling: a maximum current of \SI{200}{\pico\ampere} can be reached during the first four cycles, while, from the fifth cycle on, the maximum currents increase by more than one order of magnitude. It can also be noticed that, while the first cycles show almost no hysteresis, from the eighth cycle on, a distinct hysteresis window opens up, bringing the domain wall to reach the lowest resistance values. This behavior suggests that Joule heating might cause the resistance changes at the domain walls. The lower current branch in Figure \ref{OOP} \textbf{c)} corresponds to the initial increase in voltage from \SI{0}{\volt} to \SI{3}{\volt}, while the higher currents are obtained for decreasing the voltage again from \SI{3}{\volt} to \SI{0}{\volt} as indicated by the arrows. This counter-clockwise hysteretic response resembles the \textit{so-called} eight-wise switching that involves interface changes, rather than formation of filaments to explain the resistive switching.\cite{Sawa2008,Cooper2017} Interestingly, the hysteresis window is opened by an abrupt current increase of over one order of magnitude, happening at different threshold voltages, whose values vary between \SI{2.5}{\volt} and \SI{3}{\volt}. The threshold voltages lack a clear trend with further cycling, which again points to Joule heating as one of the drivers of the resistance change.\par

In Figure \ref{OOP} \textbf{d)}, a domain wall \textit{I}-\textit{V} loop measured at the same location as in Figure \ref{OOP} \textbf{c)} is fitted by a model\cite{Yang2008}, which is based on a memristor and rectifier equivalent circuit, as reported for TiO$_2$ memristors. and described by:
\begin{equation}\label{eq:1}
    I(V)=w^n \beta \ \text{sinh}(\alpha V) + \chi(\text{exp}(\gamma V)-1)
\end{equation}

\noindent
where the parameters and their physical significance are described in the Supporting Information. Despite obvious differences of our material compared to the TiO$_2$ memristors,\cite{Yang2008} the model captures the main shape of the \textit{I}-\textit{V} loop and provides a quantitative functional expression for the simulation of the OOP memristive behaviour of individual DWs, which is an important prerequisite for the design of circuits that incorporate these DWs. More information about the model and the results o the fits can be found in the Supporting Information. \par

\par

\begin{figure}[htbp]
        \centering
		\includegraphics[draft=false,width=\textwidth]{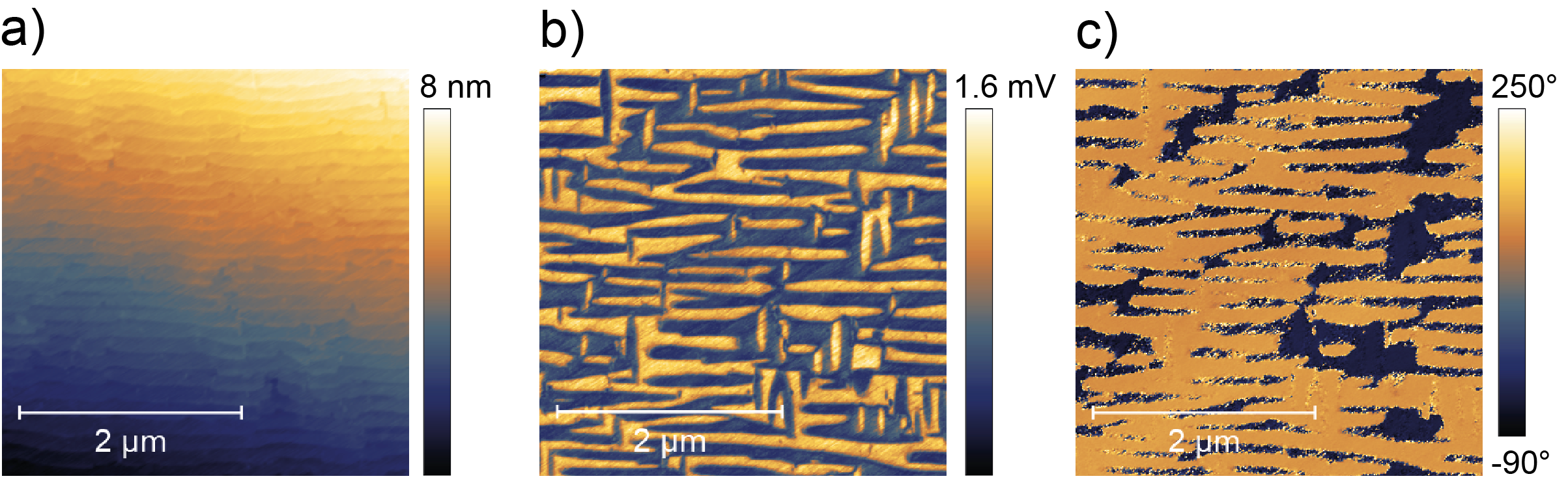}
 \caption[]{\textbf{a)} Topography, \textbf{b)} lateral PFM amplitude and \textbf{c)} lateral PFM phase of a BFO/STO sample with a \SI{55}{\nano\meter} thick BFO layer. All images show the same region of the sample.}
	\label{All_IP_PFM}
\end{figure}
\noindent

As IP conductivity and connectivity of the DW network can only be investigated in the absence of a bottom electrode, these measurements are performed on the BFO/STO samples. In Figure \ref{All_IP_PFM} \textbf{a)}, the BFO topography indicates a high-quality BFO layer with atomically flat terraces, which result from epitaxial growth on the terrace-step structure of the bare STO substrate. The lateral PFM amplitude and phase signals on the same BFO/STO sample are shown in Figure \ref{All_IP_PFM} \textbf{b)} and \textbf{c)}, respectively. The ferroelectric domain structure is of the same type as that observed in the BFO/SRO/STO samples (i.e. four types of \SI{71}{\degree} down-polarized domains are present). However, due to the absence of the SRO buffer layer, the domain morphology of this sample differs from that of the BFO/SRO/STO sample shown in Figure S1 in that the BFO/STO sample exhibits longer stripe domains interrupted by smaller orthogonal domains. The domains show a well defined periodicity of approximately \SI{190}{\nano\meter}, which agrees with the width of the substrate terraces. Indeed, by comparing the topography in Figure \ref{All_IP_PFM} \textbf{a)} with the PFM images in Figures \ref{All_IP_PFM} \textbf{b)} and \textbf{c)}, a strong preference exists for the long stripe domain walls to align with the step edges. This is also directly visualized in Figure S2 (see Supporting Information).

 Most long stripe domains even nucleate directly at the edge of the step terraces, as earlier reported.\cite{Chen2007,Chu2007} Therefore, it is possible to tune the periodicity and the configuration of the domain wall network, to some extent, by changing the substrate miscut (terrace width) and the orientation of the miscut (surface plane) with respect to the crystallographic planes.

\begin{figure}[htbp]
        \centering
		\includegraphics[draft=false,width=\textwidth]{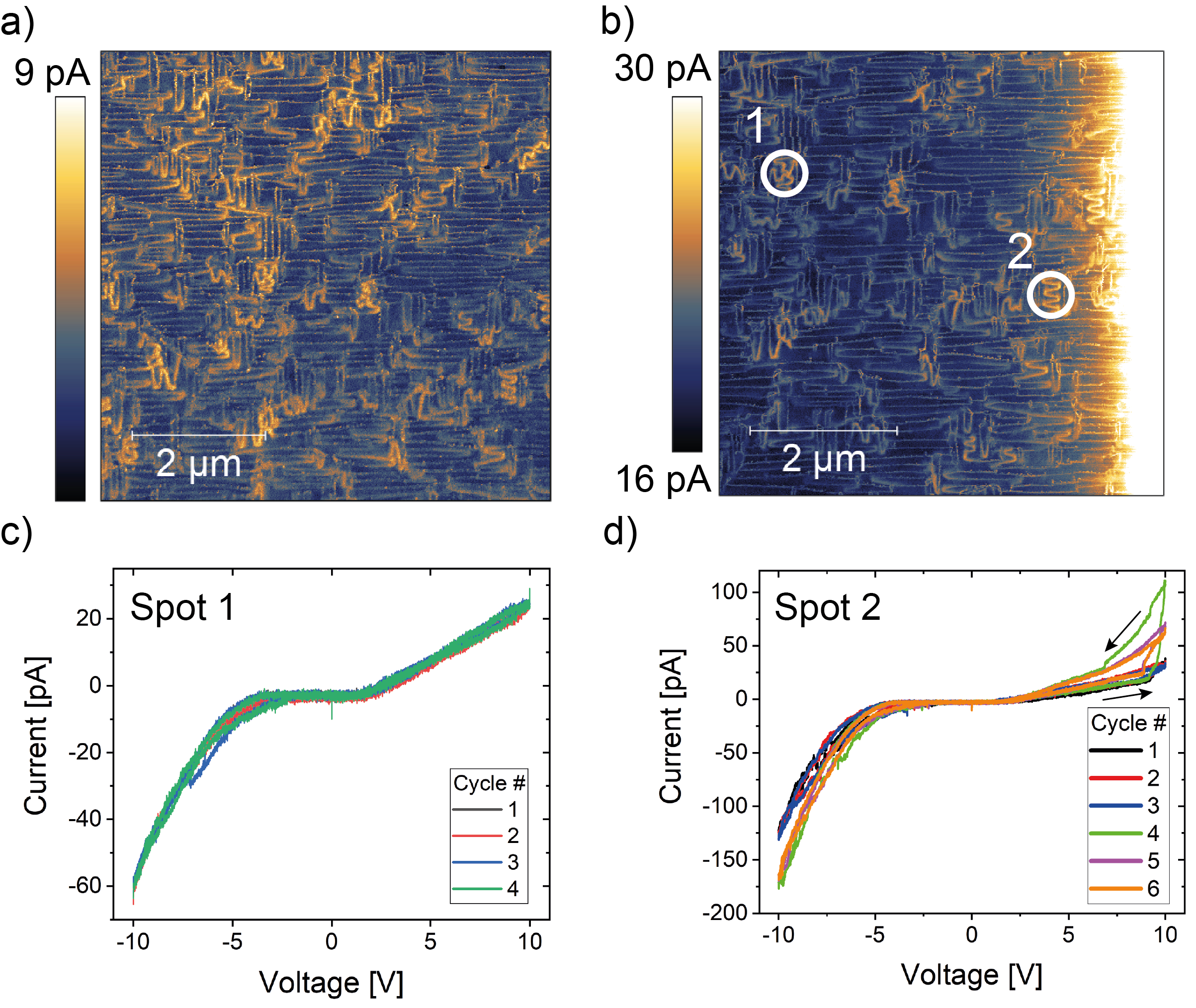}
	\caption{IP domain wall conduction. cAFM measurements on a BFO/STO sample with a \SI{55}{\nano\meter} thick BFO layer, measured in IP geometry. In \textbf{a)} the right border of the scanned area is parallel to the Pt electrode edge at an approximate distance of \SI{0.5}{\micro\meter} and a \SI{4}{\volt} sample bias is applied; in \textbf{b)} the scanned area is in the immediate proximity of the Pt electrode (visible on the right side of the map) and the sample bias is \SI{3.5}{\volt}. The color scale maximum of \SI{30}{\pico\ampere} was chosen to improve visibility.	The location of conduction map \textbf{b)} is offset by approximately \SI{0.5}{\micro\meter} to the right and \SI{0.5}{\micro\meter} to the bottom with respect to that in \textbf{a)}. \textit{I}-\textit{V} sweeps on two DWs are shown in \textbf{c)} and \textbf{d)}. Their approximate locations are given by the circles in \textbf{b)} labelled as Spot 1 (\textbf{c)}) and Spot 2 (\textbf{d)}).}
\label{IP}
\end{figure}

A conduction map of the BFO/STO sample obtained under the application of a \SI{4}{\volt} sample bias is shown in Figure \ref{IP} \textbf{a)}. The right border of the scanned area is parallel to the right edge of the Pt electrode window and approximately \SI{0.5}{\micro\meter} away from it. The location of the scanned area with regard to the window in the Pt electrode is shown in Figure S3 \textbf{a)} (see Supporting Information). Other scanned areas present similar DW structure and current levels also at longer distance (several \SI{10}{\micro\meter}) from the Pt electrode edge as depicted in Figure S3 \textbf{b)} and \textbf{c)} of the Supporting Information. As in the case of the BFO/SRO/STO samples, the conductivity is clearly enhanced at the domain walls and the long-stripe domain structure is also visible in the conduction maps. Compared to the OOP conductivity maps, the observed currents are strongly reduced, as expected in this configuration, with the electrodes being further apart. Typical currents in the long DWs are between \SI{2.5}{\pico\ampere} and \SI{4}{\pico\ampere}, while the shorter, wiggling DWs exhibit currents between \SI{5.0}{\pico\ampere} and \SI{7.0}{\pico\ampere}. The latter DW type also displays a larger apparent domain wall width. Figure \ref{IP} \textbf{b)} shows another conduction map of the same sample directly measured next to the edge of the Pt electrode. In this scan the applied bias is reduced to \SI{3.5}{\volt} (by close inspection, the overlapping area between the two maps can be recognized). \par

\noindent

From Figures \ref{IP} \textbf{a)}-\textbf{(b)}, a horizontal current gradient is only visible in the close proximity of the electrode edge. For a tip-Pt electrode distance larger than \SI{2}{\micro\meter} almost no current gradient is visible. The distribution of the electric field in the BFO/STO sample, for a bias of \SI{3.5}{\volt}, can be simulated using finite element methods (FEM), as shown in the Supporting Information (see Figure S4). Two types of simulations are performed. First, the electric field close to the Pt electrode is simulated (Figure S4 \textbf{a)}). Due to the edge effect at the \SI{20}{\nano\meter} thick Pt electrode, the magnitude of the electric field sharply decreases by about 70\% over only \SI{10}{\nano\meter} distance to the electrode edge. This is in quantitative agreement with the observations, and we can, thus, state that the observed current gradient in Figure \ref{IP} \textbf{b)} is directly linked to the steep electric field gradient at the electrode edge. Second, the electric field distribution around the tip is also simulated to explain the absence of a current gradient beyond this edge effect. Such gradient is expected for an effective resistance governed by the length of the domain walls. In Figure S4 \textbf{b)} and \textbf{c)} (see Supporting Information), a FEM simulation of the IP conduction sample, including the microscope tip, is shown. For the purpose of estimating the field distribution, the BFO thin film is considered to be homogeneous (i.e. the simulation does not contain conducting DWs). The FEM shows that the potential difference is largely enhanced close to the tip, indicating that the observed conductivity corresponds to a strongly localized area around the tip, with the contribution from other areas of the network being negligible. This explains the absence of the gradient across the IP conduction maps. 

However, it is also important to notice that, even in the case of a stand-alone device with extended electrodes and a well-defined (homogeneous) potential difference across a memristor/resistor network, a smooth decrease of the current at increasing distances from the electrode is not necessarily expected. To show this, the notion of effective resistance on graphs is used. The effective resistance on a graph is defined as a distance measured between a pair of nodes, by viewing the graph as an electric circuit with a \SI{1}{\volt} voltage source connected between the selected nodes. Each edge in the network and its corresponding weight are associated to a resistor and its resistance value, respectively.\cite{Ellens2011}

\begin{figure}[htbp]
        \centering
		\includegraphics[draft=false,width=0.7\textwidth]{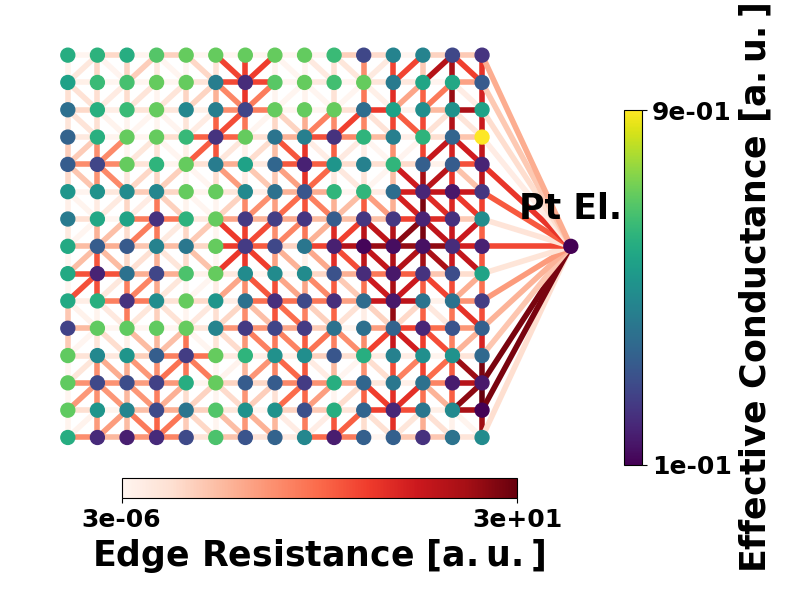}
	\caption{Resistor network showing that an evenly distributed gradient from the Pt electrode is not necessarily expected. The red color of the edges is proportional to their resistance. The node color is proportional to the effective conductance between the node and the reference Pt electrode on the right of the network. The values are in arbitrary units (a.u).}
\label{gridgraph}
\end{figure}

With that purpose, the conduction map in Figure \ref{IP} \textbf{a)} was coarse-grained into $15\times15$ patches and pre-processed to enhance the contrast in order to emphasize the lack of an evenly distributed gradient (see Figure S5 \textbf{a)}-\textbf{c)} of the Supporting Information). Then, a grid-graph resistor network with a number of nodes equal to the number of patches was built. Additional diagonal edges were then also included in this lattice. One among those edges (randomly chosen for each node) was excluded to avoid local intersections of the edges, in order to maintain a two-dimensional framework.\cite{Montano2022} Finally, a reference node representing the Pt electrode was included on the right-hand side of the network. The resistivity of the edges in the circuit graph was then fitted to reproduce the same effective conductance distribution on the graph as that obtained from the processed conduction map.\cite{SciPyProceedings_11,Virtanen2020} Figure \ref{gridgraph} shows that it is possible to obtain a distribution of effective conductance between each node/tip position and the Pt electrode node that does not show the evenly distributed gradient expected for increasing Euclidean distances from the Pt electrode in case of isotropic conductance. Thus, this modeling result highlights the role of the underlying network structure on the effective conductance distribution over the nodes. 

\noindent

For the BFO/STO sample measured in Figure \ref{IP} \textbf{b)}, also local \textit{I}-\textit{V} sweeps on DWs are performed. In Figure \ref{IP} \textbf{c)}, the \textit{I}-\textit{V} curves obtained from triangular sweeps with \SI{1}{\hertz} frequency and \SI{10}{\volt} amplitude on a DW are shown. The location of the measured DW is marked by the circle labelled as "Spot 1" in Figure \ref{IP} \textbf{b)}. The tip-electrode distance is approximately \SI{5}{\micro\meter}. The \textit{I}-\textit{V} characteristics are asymmetric, but clearly different from the rectifying behaviour observed in the OOP measurements shown in \ref{OOP} \textbf{b)}: while for positive voltages, a linear current increase up to \SI{20}{\pico\ampere} is found, the branch of negative voltages increases faster and reaches up to \SI{-60}{\pico\ampere}. The \textit{I}-\textit{V} curve of every cycle looks similar and only a weak hysteretic behavior is found.\par

Using the same triangular waveform, \textit{I}-\textit{V} sweeps are applied to a DW at a different location, which is labelled as "Spot 2" in Figure \ref{IP} \textbf{b)}. The tip-electrode distance for this DW location is approximately \SI{1.5}{\micro\meter}. For the first 3 cycles, a small hysteresis window is found. Similar to the observation of Figure \ref{IP} \textbf{c)}, ohmic-like behavior is found for positive voltages, reaching up to \SI{40}{\pico\ampere}, while the negative voltage branch displays a faster increase and reaches up to \SI{-130}{\pico\ampere}. The fourth cycle shows the largest hysteresis window with an abrupt current increase at a threshold voltage of \textit{V}$_{\text{th}}$ $\sim $ \SI{9.5}{\volt}, leading to an increased maximum current of more than \SI{100}{\pico\ampere}. From this cycle on, the maximum negative current for negative voltages is also increased to \SI{-170}{\pico\ampere}. Cycles 5 and 6 show a smaller hysteresis window accompanied with a less abrupt switching and intermediate current values for positive voltages.\par

\section*{Discussion}

As mentioned above, the complex distribution of connectivity of the DW network structure makes it challenging to intuitively predict the current distribution. As seen in Figure \ref{IP} \textbf{a)} and \textbf{b)}, the DW current of the long, quasi-periodic, DWs is lower than the current measured at the shorter wiggling DWs. This is the most visible DW current contrast in both conduction maps. One possible reason for this difference is a higher level of connectivity to the rest of the network for the wiggling DWs. All long DWs are oriented parallel to each other (horizontally, in the maps), so they display limited connectivity to the rest of the network. The wiggling DWs interrupt and connect to the long DWs, while being oriented in almost all possible directions, collecting the charge flow from a number of conduction paths at their location.\par 

Moreover, the lack of current gradient in the conduction maps, which seemed puzzling at first, is shown to arise from the measurement set-up, limiting the sensitivity to the local environment of the tip. Because of this limitation, it is important to ensure that the charge flow solely occurs laterally across the DW network and does not leak through other parts of the layer stack. 

A possible reason for such an alternative current path could be the presence of an unexpected conductive layer at the STO/BFO interface, for instance through the formation of a two-dimensional electron gas (2DEG).\cite{Ohtomo2004} The conductive interface would shunt the DW network, such that at each tip position the current flows first vertically through the DW to reach the conductive interface and then laterally through the interface towards the DWs, which are vertically connected to the Pt electrode edge. Assuming the interface conductivity being much larger than the DW conductivity, a horizontal current gradient in the conduction map would also be suppressed. 

The microscopic origin of a 2DEG could be explained to a large extent by polar discontinuities depending on the formal valence states of the crystal sublattices.\cite{Ohtomo2004} In fact, the existence of a 2DEG at the STO/BFO interface has been shown.\cite{Chen2015} While details about the STO substrate termination are not given, the authors report Ti diffusion across the interface, which is most likely a result of a TiO$_2$-terminated STO substrate. However, in the present work, the STO substrates of the BFO/STO samples exhibit SrO-termination as explained in the Methods section. The epitaxial growth of BFO on a SrO-terminated STO substrate results in a (FeO$_2$)$^-$/(SrO)$^0$ interface. Theoretically, this interface should host a two-dimensional hole gas (2DHG), as p-type carriers are attracted by the remaining negative charge. However, so far 2DHGs have been much less observed compared to 2DEGs in different material systems. Since many of these materials are oxides, typical defects such as positively charged oxygen vacancies are likely to move to the interface, neutralizing the excess negative charge and impeding the accumulation of holes. In that sense, one can argue that the (FeO$_2$)$^-$/(SrO)$^0$ interface does not show a 2DHG since the more prominent but electronically equivalent (AlO$_2$)$^-$/(SrO)$^0$ interface does not show a 2DHG either, but insulating behavior instead.\cite{Ohtomo2004} Up to now, a 2DHG has only been proven in one material system with high experimental effort to achieve extremely low oxygen vacancy densities. \cite{Lee2018a}

Thus, the presence of a conducting interface is unlikely. This is consistent with the fact that attempts to perform OOP conducting measurements in the BFO/STO samples by contacting the interface (using wire-bonding, in the same manner the SRO electrode is accessed in the BFO/SRO/STO samples), have not been successful. It is also consistent with the values of the voltages needed to induce observable currents in the \textit{I}-\textit{V} sweeps in both geometries. For a maximum voltage amplitude of \SI{3}{\volt}, the OOP currents (see Figure \ref{OOP}) are up to two orders of magnitude higher than the measured IP currents (see Figure \ref{IP}), even though a larger voltage amplitude of \SI{10}{\volt} was applied to the latter. This can be explained by the very different distances between the tip and the electrodes. This distance is equal to the BFO layer thickness of \SI{55}{\nano\meter} in the OOP maps; while typical tip-electrode distances for the IP conduction \textit{I}-\textit{V} sweeps are in the \SI{}{\micro\meter} range, leading to lower effective electric fields. 

\section*{Conclusions}

Conducting atomic force microscopy (cAFM) has been used to characterize electronic transport through domain wall (DW) networks on BiFe$\text{O}_3$ (BFO) thin films grown on STO with and without bottom electrode. This allows to compare the out-of-plane (OOP) and in-plane (IP) conductivity and characterize both the individual DWs and the network connectivity. Local current-voltage (\textit{I}-\textit{V}) sweeps in the OOP response of individual DWs show hysteresis, which can be taken as a proof of resistive switching and memristive behavior. While OOP enhanced conductivity in as-grown BFO domain walls and resistive switching behaviour of individual (artificially written) domain walls in BFO and other materials have been previously reported, the difficulties of fabricating good quality thin films without bottom electrode has made the demonstration of lateral conductivity very challenging. In this work, memristive behavior of DWs at high enough currents is observed, indicating that it originates in the local Joule heating that may induce local changes in the (defects) chemistry, as in several other memristive oxides. The memristive behaviour evolves with multiple voltage cycles at the same DW location, thus indicating plasticity of the DW network both for OOP and IP DW conduction. These results offer insight into using single DWs and DW networks for memory and neuromorphic applications, which is not only limited to BFO but can be generalized to other ferroic oxides.\par

\section*{Methods Section}

Prior to deposition of the thin films, commercially available STO (100) substrates were etched in buffered hydrofluoric acid and annealed to obtain TiO$_2$-termination and atomically smooth terraces with step edges of one unit cell \cite{Koster1998b}. For OOP conduction, a \SI{6}{\nano\meter}-thick SrRuO$_3$ (SRO) layer was deposited using PLD in an oxygen atmosphere of \SI{0.14}{\milli\bar} at \SI{610}{\celsius} prior to the growth of BFO, to serve as a bottom electrode. A \SI{55}{\nano\meter} thick BFO layer was deposited at \SI{0.3}{\milli\bar} oxygen atmosphere at \SI{640}{\celsius} substrate temperature. Both layers are successively deposited to preserve the quality of the interface. For both depositions a laser fluence of \SI{2.34}{\joule \per \cm \squared} was used. The bottom electrode was electrically contacted to the sample bias terminal of the AFM by wire-bonding. A sketch of the OOP sample layout and measurement design is shown in Figure \ref{double_setup} \textbf{a)}.\par

For IP conduction, the layer stack lacks the bottom SRO layer and a different strategy is used. Nevertheless, care has to be taken to assure the same quality of the BFO layer. High-quality epitaxial BFO films require an underlying A-site terminated layer to grow smoothly with pronounced terrace formation \cite{Solmaz2016}. In the case of OOP samples, the SRO bottom electrode layer is automatically A-site (SrO) terminated due to the high volatility of RuO$_2$ \cite{Rijnders2004}. However, for the IP samples, the termination of the STO substrate was changed from TiO$_2$ to SrO by depositing a SrO monolayer using PLD. The SrO target was produced in a solid-state synthesis from commercial SrO powder going through several steps of drying, pressing and sintering. Due to the high reactivity of SrO with $\text{H}_2\text{O}$, it is crucial to keep the SrO target from humid ambient atmosphere while handling it. The SrO layer was deposited at an oxygen atmosphere of \SI{1e-5}{\milli\bar}, \SI{850}{\celsius} substrate temperature, and a laser fluence of \SI{1.17}{\joule\per\cm\squared}. During SrO growth, the thickness was precisely controlled by RHEED such that exactly one monolayer was deposited\cite{Nie2014a,Hensling2018a}. The growth parameters for the subsequent BFO deposition were the same as for the BFO deposited for the OOP conduction samples. For all depositions a laser frequency of \SI{1}{\hertz} was used. The heating rate was \SI{30}{\celsius\per\min}, while the sample was cooled down at a rate of \SI{7}{\celsius\per\min} in an oxygen atmosphere of \SI{200}{\milli\bar}. To perform lateral conduction measurements, a \SI{20}{\nano\meter} thick Pt layer was evaporated. The Pt electrode was patterned with UV-lithography to create windows with sizes of 200 x \SI{200}{\micro\meter\squared} and 100 x \SI{100}{\micro\meter\squared}, through which the BFO can be contacted with an conducting AFM tip to perform cAFM measurements. The Pt electrode was wirebonded and connected to the sample bias terminal of the AFM, while the conducting tip was electrically grounded as can be seen in Figure \ref{double_setup} \textbf{a)}.\par

Prior to all scanning probe microscopy (SPM)  measurements the samples were cleaned using a Fischione Instruments Model 1020 Plasma Cleaner for \SI{8}{\min} with a 75\% Ar / 25\% O$_\text{2}$ gas mixture to remove any carbon-containing contamination. All SPM measurements of this work were performed in an Asylum Research Cypher ES AFM. Just before the measurement the sample was heated up in the microscope gas cell to \SI{150}{\celsius} for \SI{15}{\min} to remove excess surface humidity. During the heating and the measurements (all performed at room temperature) the gas cell was constantly flushed with Ar to provide a dry and inert atmosphere. The SPM measurements were performed using Sb-doped Si tips with a conducting CoCr coating.

Prior to the cAFM measurements, the PFM maps were obtained to reveal the ferroelectric-ferrolastic domain structure, using a similar experimental configuration as shown in Figure \ref{double_setup} \textbf{a)}.

In the cAFM setup, the sample bias was applied to the SRO bottom electrode (lateral Pt electrode), while the metallic tip was electrically grounded to perform OOP (IP) conduction measurements as shown in Figure \ref{double_setup} \textbf{a)} and \textbf{b)}. In OOP and IP cAFM, two types of measurements were performed: conduction maps that collect the current across the sample under a given bias voltage, and current \textit{versus} voltage curves collected at a fixed location on the sample, which was determined from the previously recorded conduction map. 

To perform the FEM simulations, the Electrostatics Interface of COMSOL Multiphysics\textsuperscript{\textregistered} (COMSOL, Stockholm, Sweden) was used. 

The Python language package NetworkX was used to build the circuit graph and to measure the effective resistance by its dedicated module.\cite{SciPyProceedings_11} The package SciPy was used to optimize the resistances of the edges.\cite{Virtanen2020}

\section*{Data Availability Statement}
The data that support the findings of this study is available in the open access repository DataverseNL at \href{https://dataverse.nl/dataset.xhtml?persistentId=doi:10.34894/OYIGPC}{https://dataverse.nl/dataset.xhtml?persistentId=doi:10.34894/OYIGPC}.

\section*{Supporting Information}

Supporting Information is available from the author. (See additional file: "Supporting Information")

\section*{Acknowledgements}

 We are grateful to Bart Besselink, Felix Hensling, Anne-Men Huijzer, Mian Li, Sigfried Karg, Felix Risch and Michael Wilkinson for useful discussions. We thank Jacob Baas and Henk Bonder for their invaluable technical support. We acknowledge funding from EU’s Horizon 2020, from the MSCA-ITN-2019 Innovative Training Networks programme "Materials for Neuromorphic Circuits" (MANIC) under the grant agreement No. 861153, as well as from the EU-H2020-RISE project "Memristive and multiferroic materials for logic units in nanoelectronics" (MELON) (No. SEP-2106565560). Financial support by the Groningen Cognitive Systems and Materials Center (CogniGron) and the Ubbo Emmius Foundation of the University of Groningen is gratefully acknowledged.

\newpage

%\bibliographystyle{MSP}
%\bibliography{My_Collection}

%\section*{Table of Contents text}

%Individual domain walls (DWs) in multiferroic BiFeO$_3$ thin films display resistive switching and memristive behaviour. These thin layers contain highly-dense networks of DWs, offering a physical system to explore the emergent functionality of nanoscaled memristive networks. However, demonstrating lateral conductivity has proven to be challenging. Here we show that connectivity through such a network is, indeed, possible. 

\newpage

%supp_information_begin

%\usepackage[figurename=Figure S]{caption}
\addto\captionsenglish{\renewcommand{\figurename}{Figure S}}

%\usepackage[backend=bibtex,style=numeric,sorting=none,doi=false,isbn=false,url=false,eprint=false]{biblatex}
%\usepackage[backend=biber,style=alphabetic,sorting=ynt]{biblatex}
%\usepackage{ulem}

%\usepackage[style=numeric-comp,backend=bibtex,sorting=none]{biblatex}
%\usepackage[numbers,sort&compress]{natbib}   
%\addbibresource{MyCollection.bib}
%\usepackage[nottoc]{tocbibind}
%\bibliographystyle{abbrvnat}

%\usepackage{bigstrut}
%\usepackage{multirow}
%\usepackage{array}
%\usepackage{tabularx}

%\usepackage{graphicx}
%\usepackage{amsmath}
%\usepackage{mathtools}
%\usepackage{siunitx}

%\setlength\parindent{20pt}

%\addto\extrasenglish

%\usepackage[autostyle]{csquotes}

%-----------------------
% Set pdf information and add title, fill in the fields
%-----------------------
%\hypersetup{ 	pdftitle = {AEM special issue paper},
%pdfauthor = {Jan Rieck}
%}

%-----------------------
% Begin document
%-----------------------
%\begin{document} %All text 
%\justifying
%\section{Ferroelastic domain walls in BiFeO$_3$ as networks of volatile memristive devices}
\pagestyle{fancy}

\rhead{\includegraphics[width=2.5cm]{vch-logo.png}}

\title{Ferroelastic domain walls in BiFeO$_3$ as memristive networks: Supporting Information}

\maketitle

% Author: Please give full first and last names for authors and include * after the name of all corresponding authors

\author{Jan Rieck*,}
\author{Davide Cipollini, }
\author{Mart Salverda, }
\author{Cynthia P. Quinteros, }
\author{Lambert R. B. Schomaker, }
\author{Beatriz Noheda*}

% Dedication

%\dedication{Optional dedication here. If no dedication is required, please leave blank}

% Affiliations: Please provide academic titles (Prof. or Dr.) for all authors where applicable, and include an institutional email address for all corresponding authors

\begin{affiliations}
J. Rieck, M. Salverda, Dr. C. P. Quinteros, Prof. Dr. B. Noheda\\
Zernike Institute for Advanced Materials, University of Groningen, Nijenborgh 4, 9747 AG Groningen, The Netherlands\\
Cognigron - Groningen Cognitive Systems and Materials Center, Nijenborgh 4, 9747 AG Groningen, The Netherlands\\

D. Cipollini, Prof. Dr. L. R. B. Schomaker \\
Bernoulli Institute for Mathematics, Computer Science and Artificial Intelligence, University of Groningen, 
Nijenborgh 9, 9747 AG Groningen, The Netherlands\\
Cognigron - Groningen Cognitive Systems and Materials Center, Nijenborgh 9, 9747 AG Groningen, The Netherlands\\

Cynthia P. Quinteros\\
ECyT-UNSAM, CONICET, Martin de Irigoyen 3100, B1650JKA, San Martín, Bs As, Argentina\\

Email Address: j.l.rieck@rug.nl
Email Address: b.noheda@rug.nl
%Email Address: d.cipollini@rug.nl
%Email Address: l.r.b.schomaker@rug.nl 

\end{affiliations}

% Keywords: Please provide a minimum of three and a maximum of seven keywords, separated by commas

% Abstract should be written in the present tense and impersonal style (i.e., avoid we), and be at most 200 words long

\section*{Supporting Information}

\subsection*{Piezo-force microscopy of a BFO/SRO/STO sample}

\setcounter{figure}{0} 
\renewcommand{\figurename}{Figure S}

\begin{figure}[htbp]
        \centering
		\includegraphics[draft=false,width=0.8\textwidth]{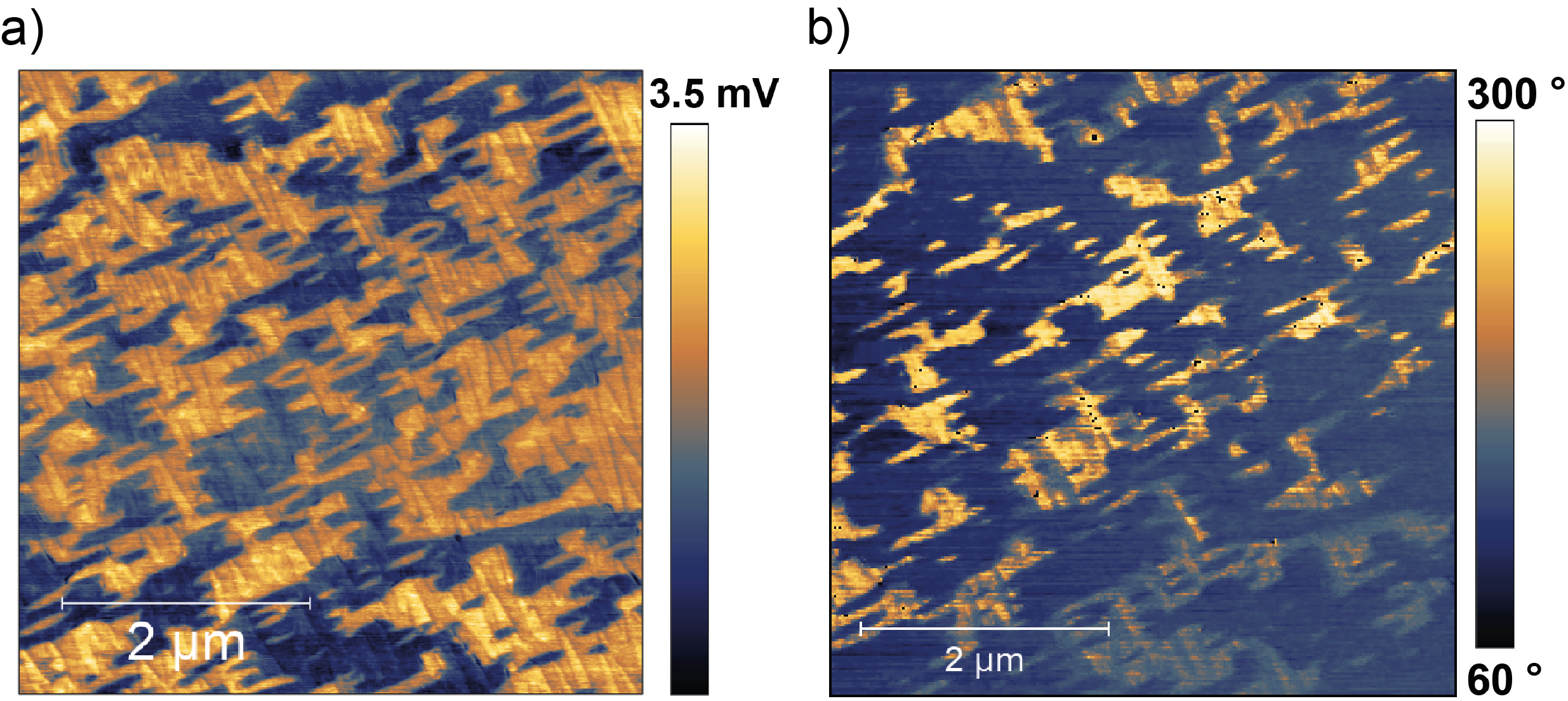}
	\caption[JR23PFM]{Piezo force microscopy (PFM) images of a BFO/SRO/STO sample showing \textbf{a)} lateral PFM amplitude and \textbf{b)} a phase  image. The contrast reflects the in-plane component of the polarization in domains separated by 71$^\text{o}$ DWs. The vertical component of the polarization points down in all the domains.}
	\label{JR23_PFM}
\end{figure}

Performing vertical PFM on the BFO/SRO/STO system produces no PFM contrast, since the as-grown BFO films are down-polarized, with only four of the eight pseudo-rhombohedral domains present.\cite{Daumont2010,Farokhipoor2011}. The lateral PFM amplitude and phase images in Figure S1 \textbf{a)} and \textbf{b)} illustrate the in-plane component of the polarization, showing the ferroelectric domain structure consistent with the presence of 71$^\text{o}$ DWs in two orthogonal directions.\cite{Daumont2010} \par

%The periodicity of the domains estimated from Figure S\ref{JR23_PFM} is approximately \SI{220}{\nano\meter}.

\subsection*{Memristor model}

The domain wall \textit{I}-\textit{V} data shown in Figure 2 \textbf{d)} is fitted by a model described by Equation 1:

\begin{equation*}
    I(V)=w^n \beta \ \text{sinh}(\alpha V) + \chi(\text{exp}(\gamma V)-1)
\end{equation*}

The sinh term in the equation represents a flux-controlled memristor in the ON state, in which the main current contribution is described by electron tunneling through a thin residual layer. $w$ is proportional to the time integral of the voltage $V$ (i.e. the magnetic flux between both terminals) and normalized to values between 0 and 1. The exponent $n$ describes the dependence of the oxygen vacancy drift velocity on the voltage applied to the device. $\alpha$ and $\beta$ are fitting constants characterizing the ON state. The exponential term represents a rectifier, modeling the current between both terminals in the OFF state with $\chi$ and $\gamma$ as additional fitting constants. More details are given in \cite{Yang2008}.\\

In Table \ref{Param_table} the fit parameters resulting from modeling the \textit{I}-\textit{V} data shown in Figure 2 \textbf{d)} with the model described by Equation 1 are given: 

\begin{table}[htbp]
  \centering

    \begin{tabular}{c|c|c|c|c|c}
    \textbf{Parameter} & n     & $\beta$ & $\alpha$ & $\chi$ & $\gamma$ \bigstrut[b]\\
    \hline
    \textbf{Value} & 10.51  & 1.35E-13 & 6.48  & 4.19E-9 & 0.04 \bigstrut[t]\\
    \end{tabular}%
      \caption{Fit parameters estimated by modeling the domain wall I-V data in Figure 2 \textbf{d)} using a memristor and rectifier model.\cite{Yang2008}}
  \label{Param_table}%
\end{table}%

\subsection*{Piezo-force microscopy of an BFO/STO sample}

In Figure S2 the topography and the lateral PFM amplitude of a 2 x \SI{2}{\micro\meter\squared} area of the BFO/STO sample for IP conduction are shown. The data was acquired in a single scan, showing alternating segments of both data channels. By comparing the neighboring topography and PFM segments, the spacial congruence of step terraces and ferroelectric domains is illustrated.

\begin{figure}[htbp]
        \centering
		\includegraphics[draft=false,width=0.4\textwidth]{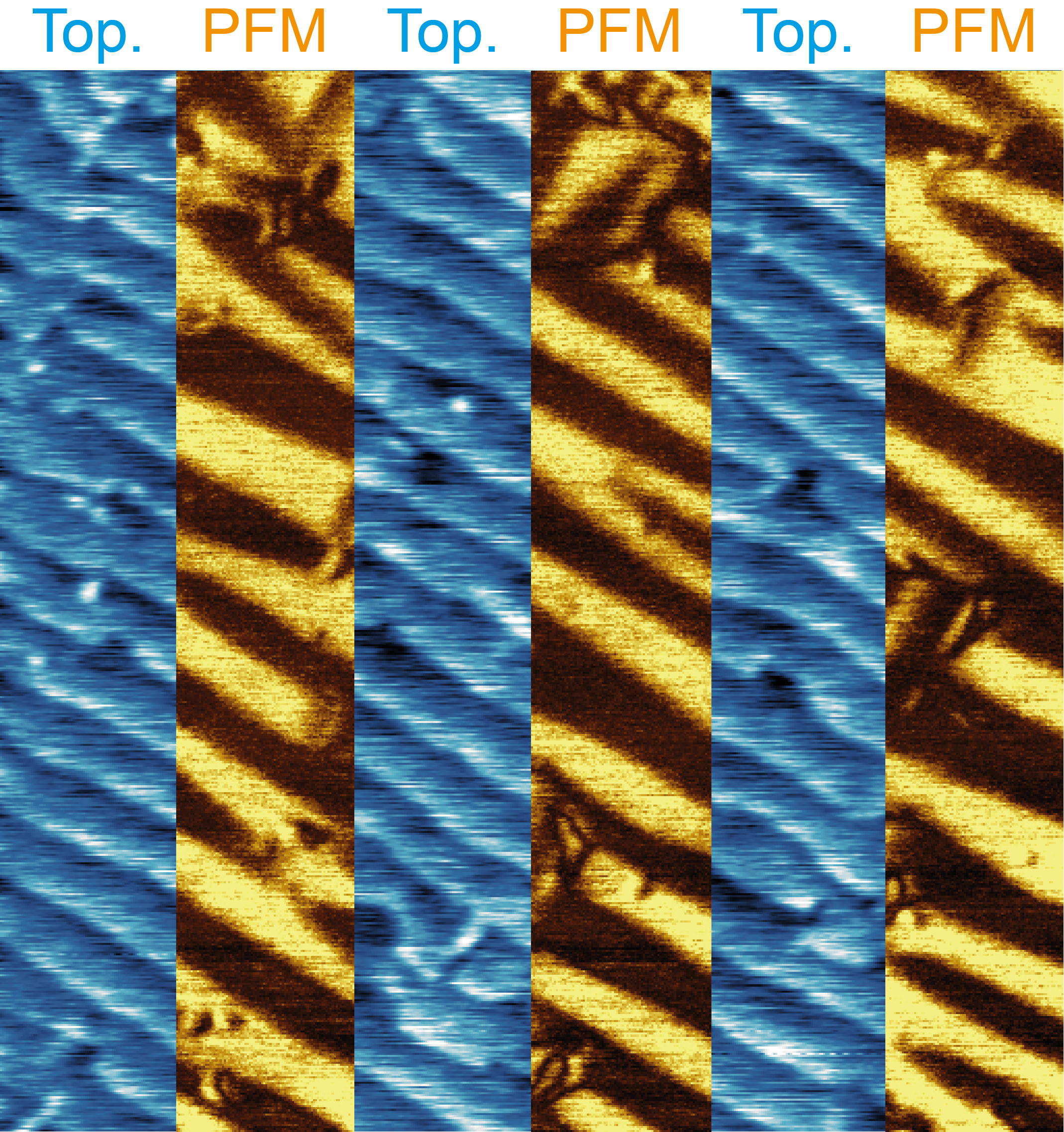}
	\caption[TopPFMcomp]{Combined topography and lateral PFM amplitude segments of the same 2 x \SI{2}{\micro\meter\squared} area measured on the BFO/STO sample to visualize the spatial coincidence of the step terraces in the topography maps with ferroelectric domains in the PFM maps.}
	\label{Top_PFM_comp}
\end{figure}

\subsection*{Additional IP conduction maps}

In Figure S3, three IP conduction maps of the BFO/STO sample of varying scan size and distance to the Pt electrode edge are depicted with the respective scan area location in the 200 x \SI{200}{\micro\meter\squared} electrode window indicated below. The conduction map of Figure S3 \textbf{a)} is also shown in the main document. In all three conduction maps, the same DW network structure consisting of long stripe DWs interrupted by shorter wiggling DWs is apparent. The maps shown in Figure S3 \textbf{a)} and \textbf{b)} are measured with the same CoCr-coated Si tip, but slightly different sample bias of \textbf{a)} \SI{4}{\volt} and \textbf{b)} \SI{3.2}{\volt}. They exhibit similar current levels reaching up to 9 and \SI{7}{\pico\ampere} respectively, even though the difference in tip - Pt electrode distance between both scan areas is larger than two orders of magnitude. The larger 13 x \SI{13}{\micro\meter\squared} conduction map shown in Figure S3 \textbf{c)} is also measured with a \SI{4}{\volt} sample bias, but a solid Pt tip and therefore exhibits higher current levels of up to \SI{28}{\pico\ampere}. Also in this IP conduction map a current gradient is absent for varying tip - Pt electrode edge distance.

\begin{figure}[htbp]
        \centering
		\includegraphics[draft=false,width=\textwidth]{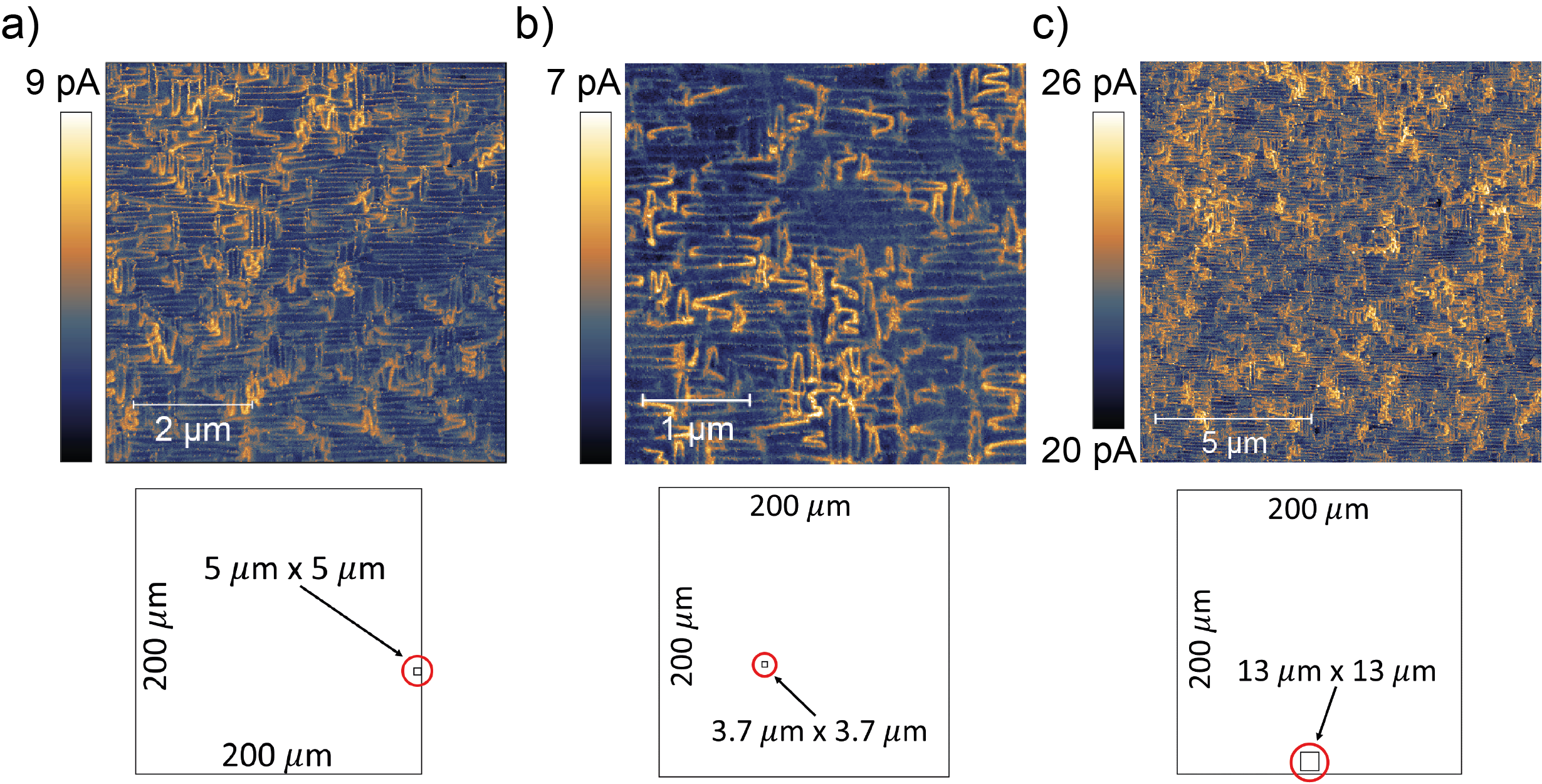}
	\caption[Latcondmapareascale]{Supplementary IP conduction maps of different sizes, tip types and locations within the 200 x \SI{200}{\micro\meter\squared} Pt electrode window. The respective scan area locations are indicated in the bottom row. The conduction maps of \textbf{a)} and \textbf{b)} are measured using a CoCr-coated Si tip and a sample bias of \textbf{a)} \SI{4}{\volt} and \textbf{b)} \SI{3.2}{\volt}. The conduction map in \textbf{c)} is recorded using a solid Pt tip and a sample bias of \SI{4}{\volt}.}
	\label{Lat_cond_maps_with_locations}
\end{figure}

\subsection*{Finite Element Modeling}

As apparent from the IP conduction maps shown in Figure S3, no current gradient for increasing tip - Pt electrode distance is found, even though it would be expected in the sense of increasing effective resistance. To explain this finding, a finite element method (FEM) simulation is performed to illustrate the distribution of electric field and potential in the BFO layer of the BFO/STO samples as shown in Figure S4. In Figure S4 \textbf{a)} a two-dimensional simulation of the electric field in the Pt electrode edge vicinity is shown. The Pt electrode is set to the typical sample bias voltage of \SI{3.5}{\volt}. The electric field is maximal directly at the Pt electrode edge, abruptly decreases within approximately \SI{10}{\nano\meter} distance and slowly goes towards 0 for larger distances from the Pt electrode edge. This electric field edge effect causes the strong current gradient visible at the right side of Figure 4 \textbf{b)}.\\

Figure S4 shows a three-dimensional Finite Element modelling (FEM) simulation of the \textbf{b)} electric potential and \textbf{c)} electric field in the BFO layer in the 200 x \SI{200}{\micro\meter\squared} electrode window including a grounded tip. The Pt electrode voltage is the same as in \textbf{a)}. Except in the direct tip vicinity, the electric potential is approximately equal to the sample bias of \SI{3.5}{\volt}. Figure S4 \textbf{c)} shows, that the electric field is always maximal and approximately of the same strength directly at the tip regardless of the tip location. Therefore, only the strongly localized area around the tip is contributing to the measured current in cAFM measurement and other, further distant sample areas can be neglected. This explains the absence of any electrode distance related gradient apart from the case shown in Figure S4 \textbf{a)}. The large difference in electric field strength of several orders of magnitude between Figure S4 \textbf{a)} and \textbf{c)} can be related to the existence of the grounded tip in the latter case.

\begin{figure}[htbp]
        \centering
		\includegraphics[draft=false,width=\textwidth]{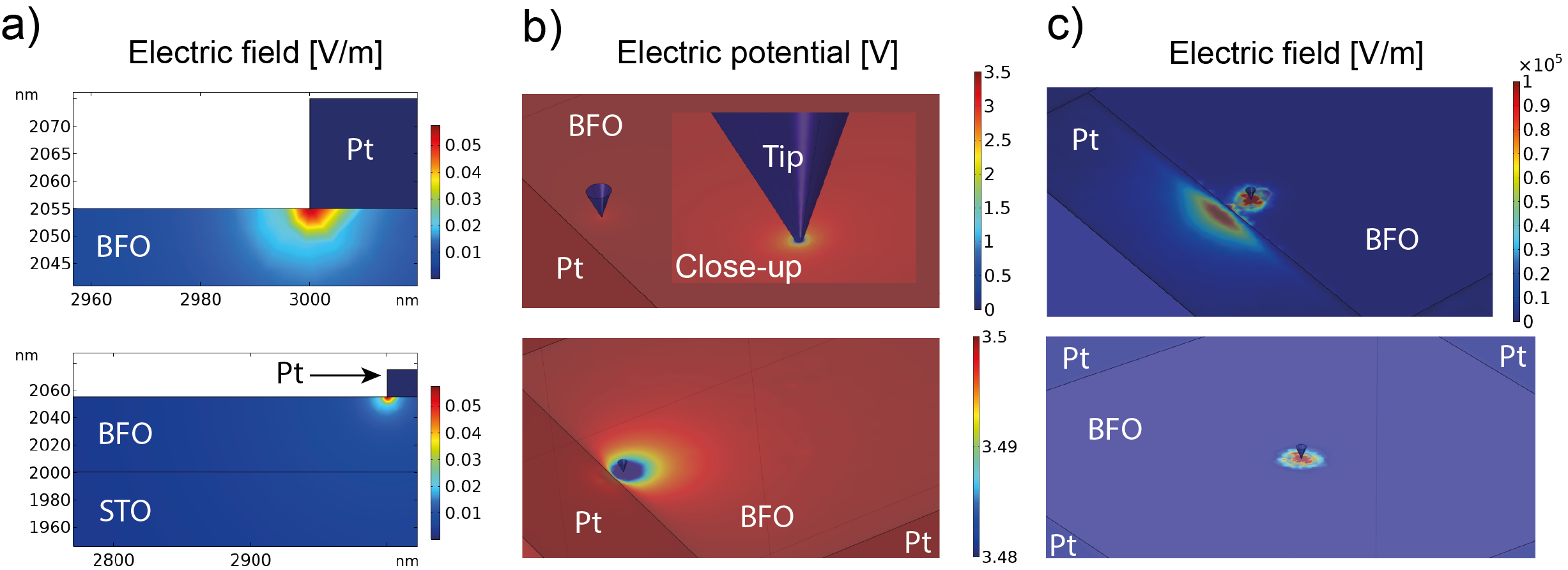}
	\caption[IP conduction COMSOL simulation]{Finite element method (FEM) simulations of electric field and electric potential in the BFO layer of the BFO/STO sample with the Pt electrode at \SI{3.5}{\volt} sample bias. \textbf{a)} Two-dimensional simulation showing the electric field gradient in direct vicinity of the Pt electrode edge. \textbf{b)} Three-dimensional simulation of the electric potential including a grounded tip with a \SI{20}{\nano\meter} tip radius. The bottom figure shows the potential in a narrow color range covering \SI{0.02}{\volt}. \textbf{c)} Electric field based on the same simulation shown in \textbf{b)} for two different tip locations: near Pt edge (top) and middle of electrode window (bottom).}
	\label{E_field_sim}
\end{figure}
\noindent

\subsection*{Graph representation of the network}

Figure S5 shows the pre-processing steps applied to the original IP conduction map from Figure 4 \textbf{a)} in order to obtain a distribution of effective conductance devoid of the expected gradient from the Pt electrode, which is shown in Figure 5.\\

\begin{figure}[htbp]
        \centering
		\includegraphics[draft=false,width=0.8\textwidth]{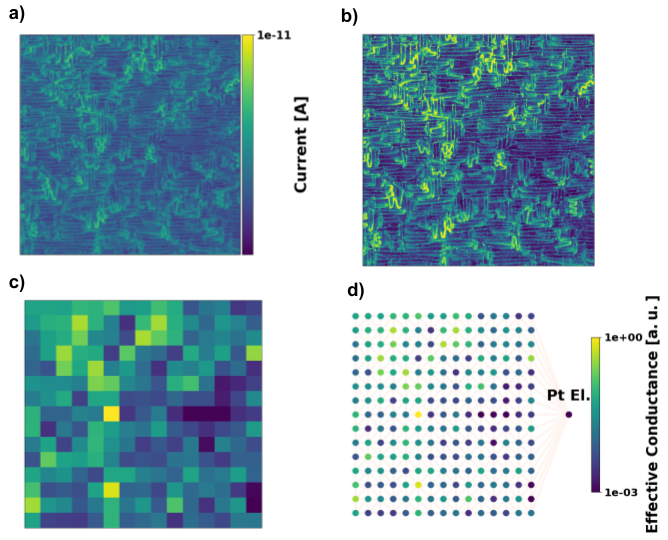}
	\caption[]{Pre-processing steps for the resistor network shown in Figure 5. In \textbf{a)} the IP conduction map from Figure 4 \textbf{a)} is shown. In \textbf{b)} a contrast enhancement is applied to emphasize the absence of a gradient from the Pt electrode on the right of the conduction map to the tip (see Figure S2 \textbf{a)}). In \textbf{c)} the current map is coarse-grained into $15 \times 15$ square patches, then the mean value over each patch is measured and finally it is scaled between the range \SI{1e-3}{} and 1. In \textbf{d)} the distribution of values obtained in the previous steps is arbitrarily defined as the effective conductance measured in a grid-graph with random diagonal edges and $15\times 15$ nodes, in which each edge is a resistor.}
	\label{preprocessing_for_gradient}
\end{figure}

\newpage

\bibliographystyle{MSP}
\bibliography{My_Collection}
%\bibliographystyle{MSP}
%\bibliography{test}

%\end{document}

%supp_information_end

\end{document}